\documentclass{optica-article}

\journal{opticajournal} 

\articletype{Research Article}

\begin{document}

\title{High-gain optical amplification and lasing from erbium-doped single-crystal films epitaxially grown on silicon}

\author{Xuejun Xu,\authormark{1,*} Tomohiro Inaba,\authormark{1} Takuma Aihara,\authormark{2} Atsushi Ishizawa,\authormark{3} Takehiko Tawara,\authormark{4} and Haruki Sanada\authormark{1}}

\address{\authormark{1}Basic Research Laboratories, NTT, Inc., 3-1 Morinosato Wakamiya, Atsugi-shi, Kanagawa 243–0198, Japan\\
\authormark{2}Device Technology Laboratories, NTT, Inc., 3-1 Morinosato Wakamiya, Atsugi-shi, Kanagawa 243–0198, Japan\\
\authormark{3}College of Industrial Technology, Nihon University, 2-1 Izumi-cho 1-chome, Narashino-shi, Chiba 275–8575, Japan\\
\authormark{4}College of Engineering, Nihon University, 1 Aza-Nakagawara, Tokusada, Tamuramachi, Koriyama-shi, Fukushima 963–8642, Japan}

\email{\authormark{*}xuejun.xu@ntt.com} 


\begin{abstract*}
On-chip erbium-doped optical amplifiers and lasers are essential for realizing fully integrated active silicon photonic circuits, but their performance has been limited by the low gain of amorphous host materials and the difficulty of direct integration on silicon. Here, we demonstrate optical amplification and lasing from erbium-doped single-crystal gadolinium oxide (Er:Gd$_2$O$_3$) thin films epitaxially grown on silicon. Optical gain measurements on waveguides fabricated on this platform exhibit a giant material gain of $78.3\pm2.1$~dB/cm and an on-chip net gain exceeding 13~dB in a 6-mm-long waveguide at 2.3~K, while a measurable gain is maintained up to room temperature. Continuous-wave lasing with low threshold, narrow linewidth, and large side-mode suppression ratio is also demonstrated in Er:Gd$_2$O$_3$ microring resonators. These results establish Er:Gd$_2$O$_3$ as the first monolithic crystalline gain medium directly integrated on silicon, providing a scalable route toward high-performance cryogenic and quantum photonic integrated circuits.
\end{abstract*}

\section{Introduction}
Erbium-doped (Er-doped) solids are key optical gain media owing to the long-lived and efficient intra-4f transitions of Er ions in the telecommunication band. Er-doped fiber amplifiers (EDFAs) have revolutionized optical communication networks by enabling low-noise, high-gain amplification of optical signals without resorting to optical–electrical conversion. Following this success, substantial efforts have focused on translating the gain functionality of Er-doped media into compact, chip-scale devices compatible with photonic integrated circuits (PICs), which will enable energy-efficient and scalable optical systems \cite{bradley2011erbium, pollnau2018optically}. On-chip Er-doped thin films capable of providing optical amplification or lasing are regarded as promising active media for monolithic silicon (Si) photonics.

Amorphous materials, including oxide glasses of various compositions~\cite{huang2002fiber, della2006compact, psaila2007er, toney2008active}, Al$_2$O$_3$~\cite{bradley2010gain, ronn2019ultra, pollnau2018optically}, Ta$_2$O$_5$~\cite{subramanian2010erbium}, TeO$_2$~\cite{frankis2020erbium}, and SiN~\cite{liu2022photonic, liu2024fully}, have been widely explored as hosts for Er ions in on-chip devices owing to their well-established and flexible fabrication processes. However, despite extensive research, their intrinsic disorder and limited compatibility with the Si photonics platform have fundamentally restricted their practical device performance. The optical gain per unit length achievable in these materials remains low, typically below a few dB/cm. As a result, amplifier and laser devices based on them require long propagation lengths, often tens of centimeters, making their footprints significantly larger than those of other integrated photonic components such as modulators and photodetectors. This fundamental limitation arises from the intrinsically small absorption and emission cross sections of Er ions in disordered matrices and from restricted doping concentrations imposed by clustering and quenching. Furthermore, these amorphous materials are generally incompatible with the silicon-on-insulator (SOI) platform widely used in Si photonics. They typically have much lower refractive indices than Si, which confines optical fields mainly within the Si layer rather than the gain medium. In addition, their performance often critically depends on achieving extremely low propagation losses that are difficult to realize on the SOI platform because of the high refractive index contrast. As a result, current Er-doped thin-film devices, though functional, suffer from large footprints and limited scalability.

Crystalline host materials have recently emerged as a promising solution to these limitations. In single-crystal matrices, Er ions occupy well-defined substitutional sites, yielding larger optical cross sections and allowing higher doping concentrations without clustering~\cite{bradley2011erbium}. Consequently, significantly higher optical gain has been demonstrated in various single-crystal systems, including KGd$_x$Lu$_y$Er$_{1-x-y}$(WO$_4$)$_2$~\cite{vazquez2018high}, Er-doped chloride silicate nanowires~\cite{sun2017giant}, and Er-doped lithium niobate on insulator (LNOI)~\cite{chen2021efficient, luo2023advances}, with reported gains exceeding 10–100~dB/cm. However, these materials face severe limitations for integration on Si: they require special crystalline substrates (KY(WO$_4$)$_2$) for high quality film growth, have difficulty being fabricated into sufficiently long waveguides to achieve usable overall gain due to the limitation of the growth method, or lack selective doping control for passive and active device integration on a single chip. Therefore, realizing Er-doped single-crystal gain media monolithically grown on Si remains an unresolved and critical challenge.

To address these challenges, we propose single-crystal rare-earth oxides (REOs) as host materials for Er ions~\cite{saini2004er2o3, kuznetsov2014single, michael2008growth, kahn2008amplification, tawara2013population, tawara2017mechanism, inaba2018epitaxial, xu2020epitaxial}. Most REOs possess cubic crystal structures with lattice constants close to twice of that of Si, allowing direct epitaxial growth on Si substrates~\cite{osten2008introducing}. Among them, gadolinium oxide (Gd$_2$O$_3$) is particularly attractive because of its extremely small lattice mismatch ($\sim$0.46\%) to Si, high chemical stability, and wide range of optical transparency from the visible to the infrared \cite{wang2009crystal, dargis2014monolithic}. Er ions can substitute for Gd ions in the lattice without forming clusters, enabling controllable doping from parts-per-million levels to fully concentrated stoichiometric Er$_2$O$_3$.

In this work, we demonstrate optical amplification and lasing from Er-doped single-crystal Gd$_2$O$_3$ thin films epitaxially grown on Si. We show that low-loss Er:Gd$_2$O$_3$ waveguides exhibit large optical gains from cryogenic to room temperatures, with a maximum material gain of $78.3 \pm 2.1$~dB/cm at 1536~nm and 2.3~K. Furthermore, we achieve optically pumped lasing in high-$Q$ microring resonators fabricated on the same platform. These results establish Er-doped single-crystal Gd$_2$O$_3$ as a promising material system for monolithic Si photonics, opening a pathway toward compact optical amplifiers, on-chip lasers, and quantum photonic integrated circuits.

\section{Material and Device Structure}
Er:Gd$_2$O$_3$ thin films with different Er doping concentrations were grown on SOI (111) substrates by molecular beam epitaxy~\cite{xu2020epitaxial, inaba2024improving}. The as-grown thin films were characterized using a series of structural and spectroscopic tools, confirming their single-crystal nature and high crystalline quality. Strong photoluminescence with a consistent spectral shape in the telecommunication band was observed from all samples with different doping concentrations. We found that the interface between Er:Gd$_2$O$_3$ and Si plays an important role in the optical properties of Er ions. Introducing a thin ($\sim$10~nm) undoped Gd$_2$O$_3$ layer to separate the Er ions from the interface significantly enhanced both the light-emission intensity and the optical lifetime. Therefore, for subsequent device fabrication and characterization, samples with a thin undoped Gd$_2$O$_3$ buffer layer were used. The growth and characterization of Er:Gd$_2$O$_3$ thin films are described in detail in Supplement 1, Sec. S1.

\begin{figure}[htbp]
\centering\includegraphics[width=\linewidth]{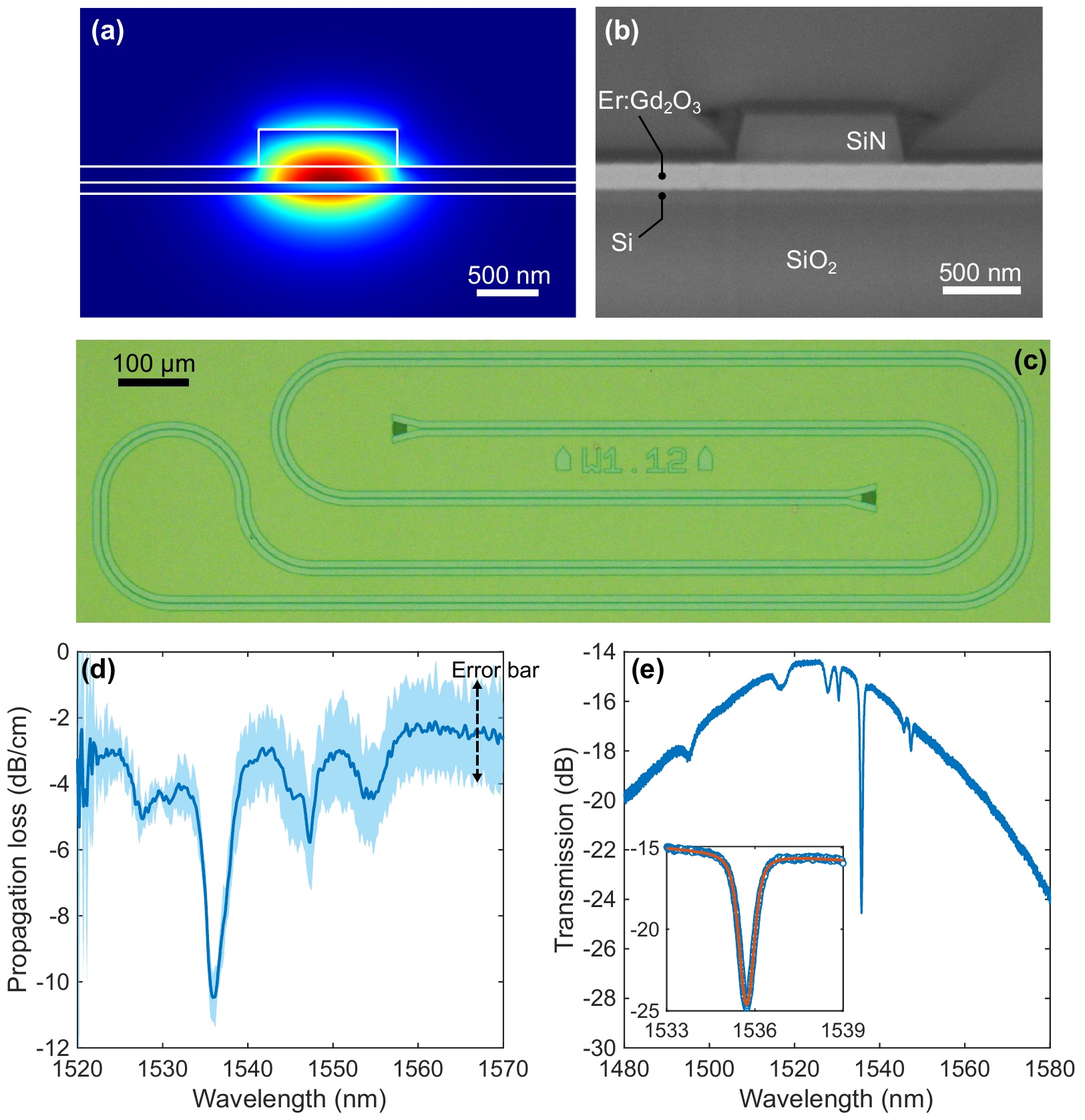}
\caption{Design and characterization of Er:Gd$_2$O$_3$ waveguides. (a) Mode profile of the fundamental TM mode of the SiN/Er:Gd$_2$O$_3$/SOI strip-loaded waveguide. The parameters of the waveguide are: Si layer thickness $t_{\mathrm{Si}} = 90$~nm, Er:Gd$_2$O$_3$ layer thickness $t_{\mathrm{Gd2O3}} = 130$~nm, SiN layer thickness $t_{\mathrm{SiN}} = 300$~nm, and waveguide width $w = 1120$~nm. (b) Cross-sectional SEM and (c) microscope images of a fabricated waveguide with input and output grating couplers. (d) Propagation-loss spectrum of the waveguide at room temperature. The light blue shaded region represents the error bars of the propagation loss for each wavelength. (e) Transmission spectrum of a 1.5-mm-long waveguide at $T = 2.3$~K, with the inset showing a zoomed view of the absorption dip around 1536~nm and the Voigt fit (solid line) to the experimental data (dots).}
\end{figure}

To demonstrate optical gain in Er:Gd$_2$O$_3$ thin films, we fabricated optical waveguide devices. A silicon nitride (SiN) strip-loaded waveguide structure~\cite{xu2021low} was designed and fabricated on samples with a Gd$_2$O$_3$ layer thickness of $\sim$130~nm and an Er doping concentration of $N_{\mathrm{Er}} = 3.13\times10^{20}$~cm$^{-3}$. The propagation loss of the waveguides strongly depends on the waveguide width, and low loss can be achieved with an appropriately chosen width, where a low-loss \textit{bound state in the continuum} is formed~\cite{zou2015guiding, yu2019photonic}. The detailed design and fabrication processes of the waveguides are described in Supplement 1, Sec. S2. Figure~1(a) shows the mode profile of the fundamental transverse-magnetic (TM) mode of the waveguide with optimized parameters, indicating a well-confined optical field. Numerical simulations using COMSOL Multiphysics indicate that approximately 35\% of the electromagnetic field energy (optical confinement factor $\Gamma_{\mathrm{Gd2O3}}$) is confined within the Er:Gd$_2$O$_3$ layer. Figure~1(b) shows a cross-sectional scanning electron microscope (SEM) image of a fabricated waveguide, in which a patterned SiN mesa is located on top of the as-grown Er:Gd$_2$O$_3$ layer on the SOI substrate. Optical waveguides with different lengths were fabricated on the same chip, one of which is shown as a microscope image in Fig.~1(c) in a spiral configuration to reduce the device footprint. A pair of focusing grating couplers was used as the input and output of the waveguide. The measured coupling efficiency of the grating couplers was approximately 26\%.

The propagation properties of the waveguides were measured at both room temperature and cryogenic temperature ($T = 2.3$~K). At room temperature, the standard cut-back method was used to determine the propagation loss of the waveguides, and the results are shown in Fig.~1(d). Several well-resolved dips appear in the spectra, corresponding to the optical transitions between the ground ($^4$I$_{15/2}$) and excited ($^4$I$_{13/2}$) states of Er ions in Gd$_2$O$_3$. Of particular importance is the deepest absorption dip at $\sim$1536~nm, which corresponds to the optical transition between the lowest Stark levels in the ground and excited states ($Z_1 \rightarrow Y_1$). The transition is known to arise from Er ions occupying the C$_2$ symmetry site, which constitutes three quarters of the Er ions. The detailed energy-level structure of Er ions in Gd$_2$O$_3$ is described in upplement 1, Fig. S3. A broadband background absorption is also present due to the overlap between broadened Stark-split transitions, but it is difficult to separate this from the passive waveguide loss. Therefore, we estimate an upper bound for the passive propagation loss of the waveguides ($\sim$2.11~dB/cm; the lowest loss across the measured wavelength range in Fig.~1(d)), which in turn provides a lower bound for the absorption due to Er ions. Considering the optical confinement factor, the peak absorption cross section of Er ions (at $\sim$1536~nm) is found to be greater than $2.31\times10^{-20}$~cm$^2$, which is indeed much larger than that of Er ions in amorphous host materials.

A typical transmission spectrum of a 1.5-mm-long waveguide measured at $T = 2.3$~K is shown in Fig.~1(e). At cryogenic temperature, several Er-related absorption dips appear with much narrower linewidths and larger depths than those at room temperature. The most pronounced dip ($\sim$1536~nm, $Z_1 \rightarrow Y_1$ transition) has an absorption linewidth of $\sim$77.6~GHz and a peak absorption cross section of $1.71\times10^{-19}$~cm$^2$, 7.4~times larger than that at room temperature. We therefore expect an even larger potential optical gain at lower temperatures, although at the expense of narrower bandwidth.

To realize lasing based on the current waveguide structure, we fabricated microring resonators. The $Q$-factor is one of the most important specifications for achieving lasing and depends on both the ring radius and the ring waveguide width, similar to the case of straight waveguides. We therefore carefully designed these two parameters to achieve high $Q$. The detailed design is described in Supplement 1, Sec. S2. Figure~2(a) shows a microscope image of a typical waveguide-coupled microring with a radius of 150~$\mu$m and an optimized waveguide width of 1.08~$\mu$m. A curved directional coupler (CDC) configuration between the bus and ring waveguides was adopted to enhance coupling efficiency [Fig.~2(b)]~\cite{hosseini2010systematic}. A typical transmission spectrum of a waveguide-coupled microring measured at $T = 2.3$~K is shown in Fig.~2(c), where periodic and sharp resonant dips are clearly seen. Most of the resonances have dip depths much larger than 10~dB, indicating operation near the critical-coupling regime. To extract the passive loaded $Q$-factor that determines the threshold gain required for lasing, we analyzed a resonance near 1538.3~nm where Er absorption is negligible. A Lorentzian fit yields a loaded passive $Q$-factor of $4.23\times10^4$, corresponding to an effective loss of 9.0~dB/cm (including both intrinsic loss of the ring waveguide and coupling-induced loss). Lasing can therefore be expected if the internal optical gain of the ring waveguide exceeds this value. Notably, the resonances span a broadband wavelength range, including pump wavelengths ($\sim$1480~nm) commonly used for Er-doped amplifiers and lasers, suggesting that resonant pumping can used to enhance pumping efficiency and lower the lasing threshold.

\begin{figure}[htbp]
\centering\includegraphics[width=\linewidth]{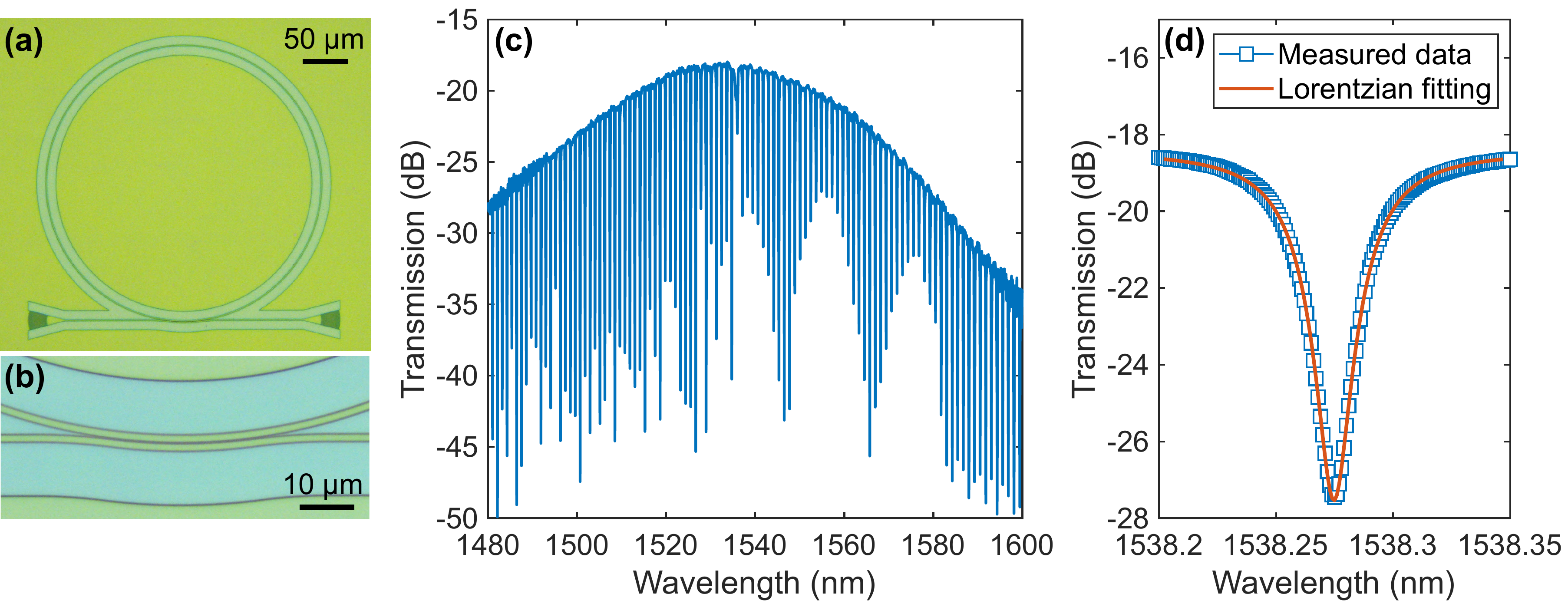}
\caption{Characterization of Er:Gd$_2$O$_3$ microring resonators. (a) Optical microscope image of a fabricated microring resonator with ring waveguide width $w_{\mathrm{ring}} = 1080$~nm, bus waveguide width $w_{\mathrm{bus}} = 1080$~nm, gap between the bus and ring waveguides $g = 300$~nm, and CDC arc length $\theta = 15^{\circ}$. (b) Zoomed-in microscope image of the CDC between bus and ring waveguides. (c) Transmission spectrum of the waveguide-coupled microring resonator at $T = 2.3$~K. (d) Zoomed-in view of a resonance near 1538.3~nm with a loaded $Q$-factor of $4.23\times10^4$ extracted through Lorentzian fitting.}
\end{figure}

\section{Optical gain in Er:Gd$_2$O$_3$ waveguides}
Pump–probe transmission measurements revealed clear optical gain in Er:Gd$_2$O$_3$ waveguides. The pump laser was a 1495~nm continuous-wave (CW) Fabry–Perot (FP) laser diode, whose wavelength is close to the $Z_1 \rightarrow Y_5$ transition of Er ions at the C$_2$ site. The measurement setup is described in detail in Supplement 1, Fig. S7. Figure~3(a) shows the transmission spectra of a 1.5-mm-long waveguide measured with and without optical pumping at $T = 2.3$~K. Several absorption dips evolved into peaks upon optical pumping, unambiguously demonstrating optical gain. We focused our subsequent measurements on the most prominent absorption dip (gain peak) around 1535.7~nm ($Y_1 \rightarrow Z_1$ transition). For the identification of the origin of the other gain peaks, please refer to Supplement 1, Fig. S8. Because the broadband background transmission outside the optical-transition wavelength range remained nearly unchanged, the material gain could be estimated directly from the gain-peak height. By fitting the gain peak at 1535.7~nm with a Voigt function [inset of Fig.~3(a)], a peak height of $\sim$4.16$\pm$0.11~dB was extracted. The peak height can be expressed as
\[
G_{\mathrm{peak}} = \Gamma_{\mathrm{Gd2O3}} g_{\mathrm{mat}} L,
\]
where $\Gamma_{\mathrm{Gd2O3}}$ is the optical confinement factor in the Er:Gd$_2$O$_3$ layer, $g_{\mathrm{mat}}$ is the material gain, and $L$ is the waveguide length. Using $\Gamma_{\mathrm{Gd2O3}} = 0.35$ and $L = 1.5$~mm for the present device, we obtained a giant material gain of $g_{\mathrm{mat}} = 78.3\pm2.1$~dB/cm.

\begin{figure}[htbp]
\centering\includegraphics[width=\linewidth]{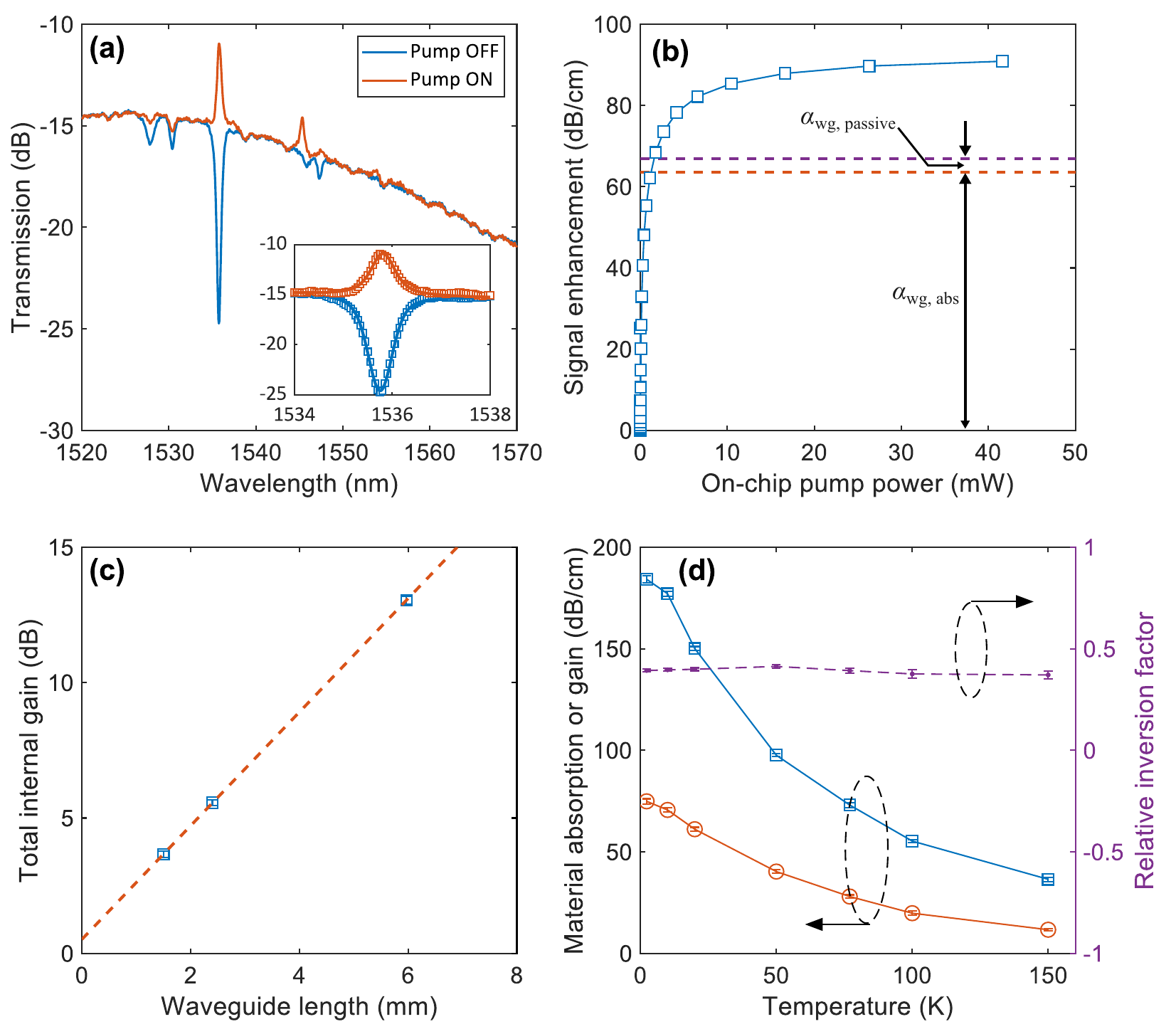}
\caption{Measurement results of optical gain in Er:Gd$_2$O$_3$ waveguides. (a) Comparison of transmission spectra of a 1.5-mm-long waveguide with and without optical pumping at $T = 2.3$~K. The pump power was 41.6~mW. (b) Dependence of optical signal enhancement on pump power. The orange dashed line represents the absorption loss of the waveguide caused by Er ions ($\alpha_{\mathrm{wg,abs}} = 63.6\pm0.8$~dB/cm), and the purple dashed line represents the background passive loss ($\alpha_{\mathrm{wg,passive}} = 3.37\pm0.03$~dB/cm). (c) Dependence of total internal optical gain on waveguide length. (d) Temperature dependence of optical absorption, gain, and relative population inversion factor of Er:Gd$_2$O$_3$ thin films.}
\end{figure}

Figure~3(b) shows the signal enhancement (SE) per unit length at 1535.7~nm as a function of on-chip pump power. The SE in dB is defined as $SE = 10 \log_{10}(T_{\mathrm{pump,on}}/T_{\mathrm{pump,off}})$, where $T_{\mathrm{pump,on}}$ and $T_{\mathrm{pump,off}}$ are the waveguide transmissions with and without pumping, respectively. As the pump power increases, the depth of the absorption dip decreases sharply. The SE in Fig.~3(b) increases to a value equal to the waveguide absorption loss ($\alpha_{\mathrm{wg,abs}}$), at which the absorption dip vanishes completely under a small pump power of $\sim$1.2~mW. When the pump power is further increased, a peak appears in the transmission spectrum. As the peak height increases, the SE surpasses the total loss ($\alpha_{\mathrm{wg,abs}} + \alpha_{\mathrm{wg,passive}}$), at which point a net optical gain is achieved. A net modal gain as large as $23.9\pm0.8$~dB/cm was obtained at the maximum pump power available in our current experimental configuration. This is among the largest modal gains demonstrated so far in Er-doped waveguide amplifiers~\cite{sun2017giant, vazquez2018high, ronn2019ultra, luo2023advances}, showcasing the great potential of our material platform for high-gain optical amplifiers. Furthermore, we measured the overall optical gain in waveguides with different lengths [Fig.~3(c)]. A total gain of 13~dB was obtained in a 6-mm-long waveguide, representing one of the shortest Er-doped amplifiers achieving >10 dB net gain. The total gain scaled almost linearly with the waveguide length without observable saturation, suggesting that even higher gains can be achieved simply by increasing the device length. The detailed spectra of waveguides with different lengths are shown in Supplement 1, Fig. S9.

We also performed gain measurements at different temperatures. The temperature dependence of material absorption and gain is shown in Fig.~3(d). Net optical gain was observed up to room temperature. However, due to significant spectral broadening of the peaks, it was difficult to extract the peak height (and thus material gain) from transmission spectra above 150~K. Therefore, data up to 150~K are plotted in Fig.~3(d). The detailed spectra at different temperatures are shown in Supplement 1, Fig. S10, together with the gain measurement results at room temperature (Supplement 1, Fig. S11) obtained in a manner similar to~\cite{ronn2019ultra}. At room temperature, the net modal gain was $1.06\pm0.77$~dB/cm; it was limited by the relatively small optical confinement factor of the waveguide, as discussed later. Both absorption and gain decreased with increasing temperature, while their ratio remained almost constant. This ratio serves as a measure of the relative population-inversion factor, defined as
\[
\eta = \frac{g_{\mathrm{mat}}}{\alpha_{\mathrm{mat}}} = \frac{\sigma_e N_e - \sigma_a N_g}{\sigma_a N_{\mathrm{Er}}} \approx \frac{N_e - N_g}{N_e + N_g},
\]
where $\alpha_{\mathrm{mat}}$ is the material absorption coefficient, $N_{\mathrm{Er}}$ is the total Er concentration, $N_e$ and $N_g$ are the population densities in the excited and ground states, respectively, and $\sigma_e$ and $\sigma_a$ are the emission and absorption cross sections. Assuming $\sigma_e \approx \sigma_a$ and negligible population in higher excited states, approximately 70\% of the Er ions are excited to the $^4$I$_{13/2}$ state in the investigated samples ($\eta \approx 40\%$).

This incomplete population inversion can be attributed to several factors. First, although we selectively excited Er ions at the C$_2$ site using a 1495~nm pump laser, part of the excited population was transferred to the C$_{3i}$ site. This can be clearly seen in the SE spectrum shown in Supplement 1, Fig. S8, where peaks corresponding to Er ions at the C$_{3i}$ site also appear. Second, for Er ions at the C$_2$ site, energy-transfer upconversion—in which Er ions are excited to higher energy levels other than $^4$I$_{13/2}$ and rapidly relax back to the $^4$I$_{15/2}$ ground state—cannot be ignored~\cite{tawara2017mechanism}. This process is confirmed by the intense green upconversion emission observed from the waveguides. Finally, quenched ions, though expected to be few, can also reduce the population inversion~\cite{agazzi2013energy}.

\section{Lasing in Er:Gd$_2$O$_3$ microring resonators}
By combining the large modal gain demonstrated in the waveguides and the high-$Q$ microring resonators, lasing can be readily achieved. For the lasing measurements, the same 1495~nm CW FP laser used for waveguide gain measurements was launched into one of the grating couplers, and the emission from the bus waveguide was collected through the other grating coupler. The measurement setup is described in detail in Supplement 1, Fig. S12. Although the linewidth of the pump laser ($\sim$1~nm) is much broader than that of the microring resonance ($\sim$0.03~nm), a fraction of the pump laser can still be coupled into the microring.

As shown in Fig.~4(a), the pump-power-dependent emission spectra around 1536~nm clearly confirm lasing. At low pump powers, a resonantly enhanced emission peak appears around 1536~nm, superimposed on a relatively broad background. Two weaker resonant peaks also appear at $\pm$1.16~nm from the central peak. At higher pump powers, both the background and side peaks are strongly suppressed, while the central peak becomes narrower and more intense, which is indicative of lasing. Figure~4(b) shows the on-chip light–light (L–L) curve and linewidth of the central emission peak as functions of on-chip pump power. The L–L curve exhibits a clear threshold at $\sim$8.5~mW, while the emission linewidth narrows as the pump power increases to $\sim$10~mW. These results provide unambiguous evidence of lasing from the microring resonators. The apparent saturation of the linewidth above $\sim$10~mW pump power is due to the wavelength-resolution limit ($\sim$20~pm) of the optical spectrum analyzer used. The actual laser linewidth is therefore expected to be much narrower than 20~pm (2.5~GHz).

\begin{figure}[htbp]
\centering\includegraphics[width=\linewidth]{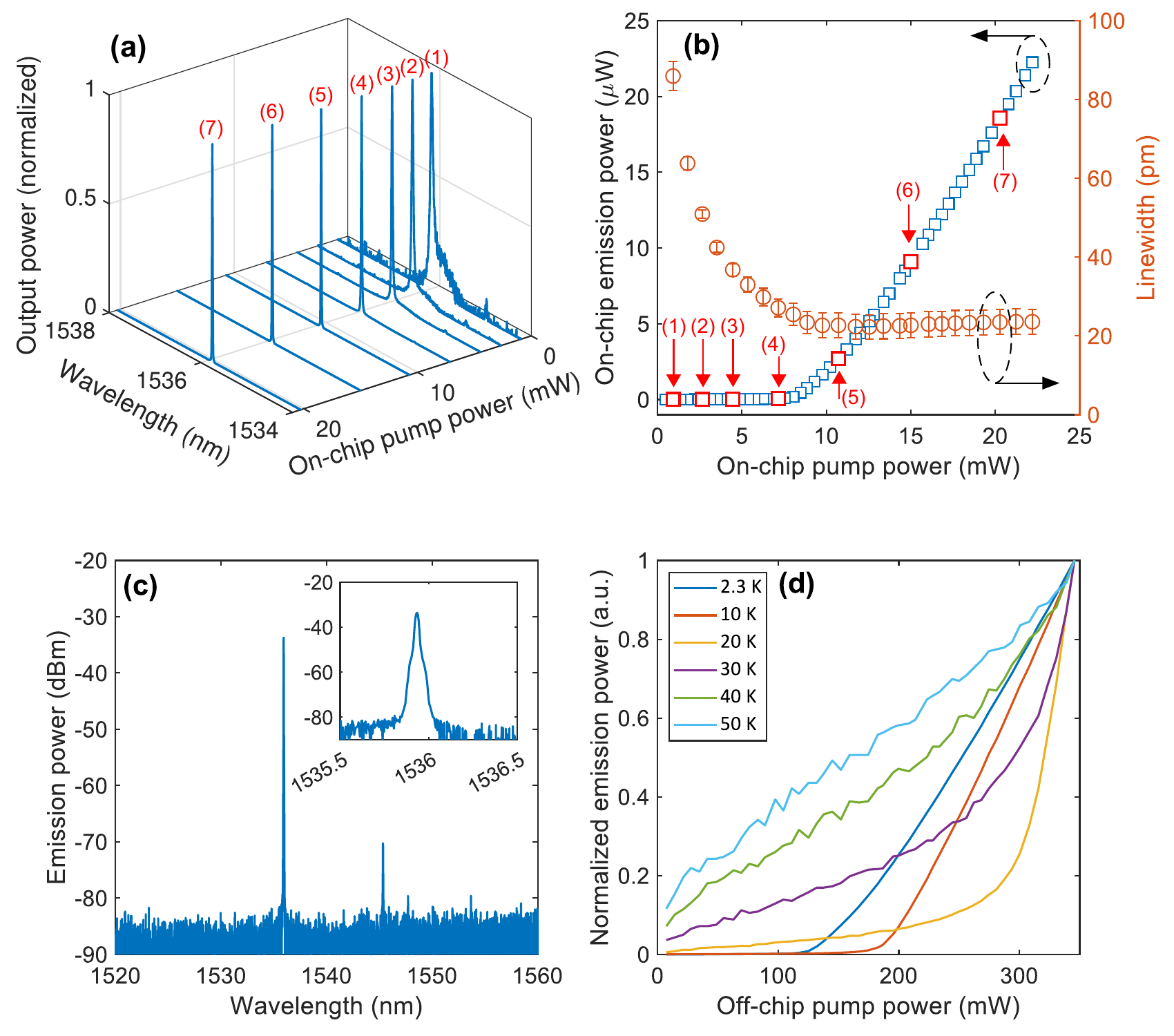}
\caption{Measurement results of lasing in Er:Gd$_2$O$_3$ microring resonators. (a) Normalized emission spectra acquired under different pump powers. The pump powers corresponding to each spectrum are indicated as red squares in (b). (b) Dependence of on-chip emission power and linewidth of the emission peak on on-chip pump power. (c) Laser emission spectrum at an on-chip pump power of 22.2~mW. The inset shows a zoomed-in view of the laser peak near 1536~nm. (d) Dependence of normalized emission power on off-chip pump power at different temperatures.}
\end{figure}

The laser spectrum over a wider wavelength range is shown in Fig.~4(c). The main laser peak at $\sim$1536~nm exhibits a large side-mode suppression ratio (SMSR) of 36.5~dB. Another weaker emission peak is around 1545.3~nm, corresponding to the $Y_1 \rightarrow Z_2$ transition of Er ions at the C$_2$ site coupled to a resonant mode; however, lasing at this wavelength was not confirmed due to insufficient optical gain. Compared with other reported microcavity lasers, such as Er-doped Al$_2$O$_3$ microring lasers~\cite{bradley2014monolithic}, silica microtoroid lasers~\cite{polman2004ultralow, yang2005erbium}, and Er-doped LNOI microring lasers~\cite{luo2021chip, yin2021electro}, the threshold pump power in our devices is relatively high. This is mainly because the pump laser used has a linewidth much broader than the microring resonance, resulting in limited pump coupling into the resonator (Supplement 1, Fig. S13). If a narrow-linewidth pump laser is used and the pump wavelength is precisely tuned to one of the resonances, a much lower threshold (in the $\mu$W range) can be expected. Another promising feature of our lasers is that they operate in true continuous-wave mode (Supplement 1, Fig. S14). No self-pulsing behavior—often caused by ion clustering in other Er-doped lasers—was observed~\cite{he2010self}. This further confirms the high crystalline quality of our Er:Gd$_2$O$_3$ thin films.

As with the optical gain measurements in waveguides, we also investigated the temperature dependence of the microring lasers. The L–L curves at different temperatures are shown in Fig.~4(d). Threshold behavior was confirmed up to $\sim$30~K, above which the emission intensity increased nearly linearly with pump power. For lasing below 30~K, the threshold pump power increases with temperature, consistent with the temperature-dependent decrease in optical gain observed in the waveguides. Based on the measured loss of the microring and the temperature dependence of the gain, lasing is expected up to $\sim$70~K (Supplement 1, Fig. S15). However, the microring resonance redshifts with increasing temperature, reducing its spectral overlap with the gain spectrum. This explains why lasing was not observed at higher temperatures.

\section{Discussion}
Our results on optical gain in waveguides and lasing in microring resonators demonstrate the great potential of single-crystal Er:Gd$_2$O$_3$ thin films for realizing high-performance, monolithic optical amplifiers and lasers on Si. At cryogenic temperatures, Er:Gd$_2$O$_3$ waveguide amplifiers exhibit outstanding performance in terms of both modal gain per unit length and total internal gain. In previously reported high-gain optical amplifiers, the net gain was either achieved only in short waveguides ($\le$1.2~mm)~\cite{ronn2019ultra} or limited by fabrication challenges that prevented long waveguides~\cite{sun2017giant}. In contrast, our Er:Gd$_2$O$_3$ platform enables long (up to 6~mm) low-loss waveguides that deliver high overall gain. As additional evidence of net optical gain, we have also demonstrated optically pumped lasing in microring resonators fabricated from the same waveguides. The lasers show excellent performance, featuring low threshold, high output power, large SMSR, and stable continuous-wave operation.

These active devices can serve as light sources for cryogenic optical interconnects~\cite{wu20212, liu2025cryogenic}, enabling high-speed and energy-efficient photonic data links between cryogenic and room temperatures~\cite{youssefi2021cryogenic, lecocq2021control}. Considering that most Si photonic device components can also operate at cryogenic temperatures~\cite{gehl2017operation, chakraborty2020cryogenic, eltes2020integrated, pintus2022integrated}, a fully monolithic photonic integrated circuit for data transfer can be envisioned. Furthermore, owing to their excellent optical and spin coherence at cryogenic temperatures, Er-doped crystals have attracted great attention as a promising platform for long-lived optical quantum memories~\cite{ranvcic2018coherence, craiciu2019nanophotonic, singh2020epitaxial, yasui2022creation, jiang2023quantum, zhang2024optical}, which are key components for long-distance quantum information networks. Narrow-linewidth lasers with frequencies matched to the optical transitions of Er ions are essential for operating such quantum memories. Our demonstrated Er:Gd$_2$O$_3$-based lasers are thus excellent candidates for this purpose. With both classical and quantum photonic devices realized on the same material system, fully integrated quantum photonic circuits become achievable.

For room-temperature applications, we identify the major obstacles to achieving higher gain and propose routes for improvement through further material and device engineering. At the current doping level, the obtained modal gain is limited by the small optical confinement factor in the Gd$_2$O$_3$ layer and the relatively large passive propagation loss. Improving these two parameters is therefore the most direct approach. The material gain at room temperature, estimated from the absorption coefficient and population-inversion factor, exceeds 9.4~dB/cm. By increasing the thickness of the Er:Gd$_2$O$_3$ film, the confinement factor can be increased to nearly 100\% (Supplement 1, Fig. S16). Advanced photonic structures such as slow-light waveguides~\cite{ek2014slow} may also be employed to further enhance the effective confinement factor. The relatively high passive propagation loss originates from deviations in waveguide dimensions and refractive index, as well as fabrication imperfections. This can be mitigated by employing novel waveguide geometries with larger tolerance to dimensional fluctuations, as we proposed recently~\cite{inaba2025a}. We expect that the propagation loss can be reduced to below 1~dB/cm. With these improvements, modal gains exceeding 8~dB/cm can be readily achieved, which is comparable to those reported in state-of-the-art Er-doped waveguide amplifiers.

A key engineering parameter that remains to be optimized is the Er doping concentration ($N_{\mathrm{Er}}$). As noted earlier, the population-inversion factor for the present doping level is approximately 40\%. Detrimental effects such as intersite energy transfer, energy-transfer upconversion, and ion quenching are expected to be alleviated at lower doping concentrations. However, a reduced $N_{\mathrm{Er}}$ will also lower the total achievable absorption and gain. A dedicated investigation into the dependence on $N_{\mathrm{Er}}$ is therefore desirable to balance this trade-off. With such optimization, our material platform could enable ultracompact on-chip optical amplifiers and lasers with device lengths orders of magnitude shorter than those of existing technologies.

\section{Conclusion}
We have demonstrated optical amplification and lasing in Er-doped single-crystal Gd$_2$O$_3$ thin films epitaxially grown on Si. Using SiN strip-loaded waveguides fabricated on this platform, we achieved unambiguous optical amplification with a giant material gain of $78.3\pm2.1$~dB/cm and an on-chip net gain exceeding 13~dB in a 6-mm-long waveguide at cryogenic temperature. Net optical gain was also observed up to room temperature. In addition, we demonstrated optically pumped continuous-wave lasing with low threshold, narrow linewidth, and large side-mode suppression ratio from microring resonators based on the same films. These results establish Er:Gd$_2$O$_3$ as the first monolithic crystalline gain medium directly integrated on Si, overcoming the intrinsic limitations of amorphous and non–Si-compatible crystalline hosts. This work provides a foundation for high-performance cryogenic optical interconnects and quantum photonic integrated circuits, offering a scalable pathway toward compact, energy-efficient, and fully integrated active devices in Si photonics.

\begin{backmatter}
\bmsection{Funding}
Japan Society for the Promotion of Science (23K26580, 23K23263, 20H00357).

\bmsection{Acknowledgment}
The authors thank Prof.~Hideki Gotoh at Hiroshima University and NTT Basic Research Laboratories, and Prof.~Junsaku Nitta at Tohoku University and NTT Basic Research Laboratories, for fruitful discussions.

\bmsection{Disclosures}
The authors declare no conflicts of interest.

\bmsection{Data Availability Statement}
Data underlying the results presented in this paper are not publicly available at this time but may be obtained from the authors upon reasonable request.

\bmsection{Supplemental document}
See Supplement~1 for supporting content.
\end{backmatter}

\bibliography{sample}

\end{document}